# Hands-on fabrication of 2D material field-effect transistors for the classroom using off-the-shelf ionic gel

*Yigit Sozen[1], Tayfun Kutlu[1], Juan J. Riquelme[1], Andres Castellanos-Gomez[1]**

[1]*2D Foundry research group. Instituto de Ciencia de Materiales de Madrid (ICMM-CSIC), Madrid, E-28049, Spain.*

*corresponding authors andres.castellanos@csic.es

## ABSTRACT

We introduce a hands-on experimental method that enables students to fabricate and operate van der Waals semiconductor-based field-effect transistors (FETs) using readily available materials and minimal infrastructure. The approach is based on ionic-gel gating using commercially available hydrogel conductors originally developed for muscle electrostimulation devices. For device fabrication we rely on commercially available thin-film interdigitated electrode arrays and for the semiconductor channel, we deposited $MoS_2$ flakes produced by high-throughput mechanical exfoliation. By avoiding the need for cleanroom processing, vacuum deposition, or custom electrolytes, this method allows functional FETs based on van der Waals materials to be realized in a standard teaching laboratory. The experiment is well suited for undergraduate and early graduate courses and is intended to complement concepts introduced in electronics and solid-state physics by providing students with first-hand experience in the fabrication and electrical characterization of emerging electronic devices. Representative electrical measurements demonstrate gate-dependent modulation of the channel conductance and illustrate characteristic features of ionic-gated transistor operation. Overall, the proposed platform provides an accessible route for students to fabricate, characterize, and analyze field-effect devices without access to conventional semiconductor foundry.

Hands-on experience in building and measuring electronic devices can provide a valuable complement to the education of scientists and engineers working in electronics, materials science, and applied physics. Accessible laboratory activities in which students can fabricate and characterize emerging electronic devices provide an opportunity for students to relate concepts introduced in lectures to experimentally measured quantities and device behaviour[1]. However, providing such practical experiences remains challenging, particularly in the context of modern transistor technologies. Conventional field-effect transistors based on complementary metal–oxide–semiconductor (CMOS) technology rely on complex fabrication processes that require cleanroom facilities, advanced lithography, high-temperature processing, and specialized equipment. Previous educational approaches have addressed transistor operation through the characterization and conceptual explanation of conventional field-effect devices[2,3], while other laboratory and training programs have provided students with hands-on experience in semiconductor processing and transistor fabrication[4–6]. Although these fabrication-oriented approaches provide valuable practical experience, they are typically implemented within specialized microfabrication laboratories. More recently, low-cost hands-on activities based on emerging transistor technologies have also been demonstrated, including the fabrication of an organic electrochemical transistor using accessible materials[7]. However, straightforward and inexpensive approaches that allow students to fabricate and characterize functioning field-effect devices remain limited. This limitation is especially significant for institutions that lack access to specialized fabrication facilities, where students may have fewer opportunities to experience both the fabrication and subsequent characterization of a functioning transistor within a single laboratory activity.

Two-dimensional (2D) semiconductors offer an attractive alternative platform for educational and demonstrative purposes[8,9]. Their atomically thin nature allows electrostatic gating to be achieved without the need for high-quality oxide dielectrics or high-temperature processing, and functional devices can be realized using minimal fabrication steps. When combined with electrolyte-based gating, large carrier-density modulation can be achieved at low voltages, further simplifying device operation and measurement. These features make 2D material field-effect transistors particularly well suited for teaching environments.

Despite these advantages, most ionic-gated transistor implementations reported in the literature rely on laboratory-developed electrolytes, such as polymer electrolytes or ionic liquids[10–13], which often require chemical preparation, controlled environments, or careful handling procedures. While simple electrolytes (e.g., aqueous salt solutions) can, in principle, be used for ionic gating, their practical use is often hindered by evaporation, limited stability, and poor reproducibility. Consequently, many studies employ engineered electrolyte systems to ensure stable operation and strong gate coupling, but these typically involve additional chemical processing steps and, in some cases, inert-atmosphere handling. Even when simplified, the handling and integration of ionic liquids or gels can be nontrivial, particularly outside specialized research environments. These factors introduce a level of complexity that is incompatible with teaching laboratories and resource-limited settings, undermining the accessibility that ionic gating could otherwise provide.

In this work, we address these limitations by demonstrating a simple and robust approach to fabricating ionic-gel-gated field-effect transistors. The method uses commercially available hydrogel ionic conductors originally developed for transcutaneous electrical nerve stimulation (TENS) devices, pre-fabricated microelectrodes, and mechanically exfoliated $MoS_2$. By relying exclusively on off-the-shelf materials and avoiding clean room processing, custom electrolyte preparation, or glovebox environments, the proposed experiment enables students to fabricate and characterize functional transistors in a standard laboratory setting. The method is particularly suitable for undergraduate and early graduate teaching, laboratory practicums, and demonstrations, showing that hands-on transistor fabrication and operation can be achieved without the need for dedicated microfabrication facilities.

## RESULTS

Figure 1a shows an image of a hydrogel ionic conductor electrode used in muscle electrostimulation and transcutaneous electrical nerve stimulation (TENS) devices. These electrodes are typically composed of polyvinyl alcohol, water, and a dissolved salt that provides the ion source. Although the exact composition is not disclosed by the manufacturer, we characterized their dielectric performance to understand their suitability for our application.

To evaluate the capacitance of the hydrogel, we prepared test capacitors by cutting square pieces of the material with areas ranging from 0.09 $cm^2$ to 0.81 $cm^2$ and sandwiched them between two metallic electrodes (aluminum foil), forming a parallel-plate geometry. The devices were then incorporated into a simple RC circuit consisting of a known series resistor $R$ and the hydrogel capacitor $C$. A square-wave voltage signal was applied, and the voltage response across the capacitor was recorded (Figure 1b).

Upon application of the voltage step, the capacitor charges and discharges exponentially with a characteristic time constant $\tau = RC$. In our case, the capacitance was extracted without fitting by determining the time required for the voltage across the capacitor to reach $1 - 1/e \approx 0.63$ of its steady-state value during the charging process (Figure 1c). This time corresponds directly to the RC time constant, allowing the capacitance to be calculated as $C = \tau/R$ . Measurements were performed at different excitation frequencies to ensure consistency of the extracted values (Figure 1d).

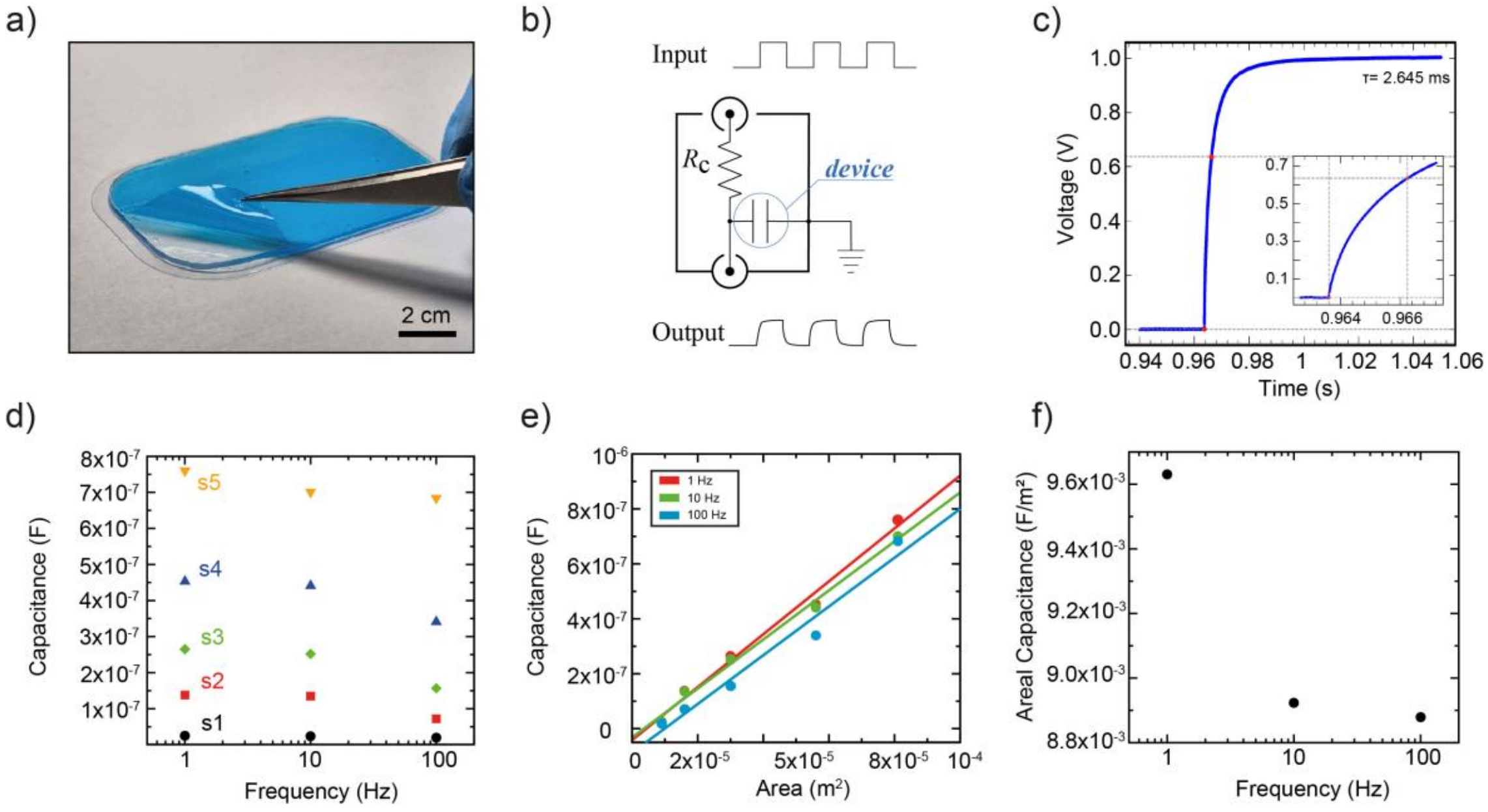

**Figure 1 | Capacitance characterization of commercially available ionic hydrogel conductors. (a)** Optical image of a square piece of hydrogel ionic conductor used in transcutaneous electrical nerve stimulation (TENS) devices. **(b)** Schematic of the RC measurement setup used to extract the capacitance. The hydrogel sample is placed between two metallic electrodes to form a capacitor connected in series with a known resistor $R_C$. A square-wave voltage input is applied, and the voltage response is recorded. **(c)** Representative charging curve of the capacitor upon application of a voltage step. The RC time constant $\tau$ is determined from the time required for the voltage to reach $1 - 1/e$ of its steady-state value, allowing extraction of the capacitance via $C = \tau/R_C$. **(d)** Extracted capacitance values for hydrogel capacitors of different areas (s1–s5) measured at different excitation frequencies. **(e)** Capacitance as a function of device area for different measurement frequencies (1, 10, and 100 Hz). Linear fits (solid lines) confirm the expected scaling with area and are used to extract the areal capacitance (capacitance per unit area). **(f)** Areal capacitance as a function of frequency, showing values on the order of $10^{-3}$ F m$^{-2}$ with weak frequency dependence in the measured range.

To improve accuracy and account for geometric variability, this procedure was repeated for capacitors with different areas. The extracted capacitance values at different frequencies (1, 10 and 100 Hz) were then plotted as a function of area, and the areal capacitance (capacitance per unit area) was obtained from a linear fit (Figure 1e). This approach reduces uncertainty associated with individual measurements and provides a more reliable estimation of the effective capacitance of the ionic hydrogel. Extracted areal capacitance values at different frequencies are given in Figure 1f. The extracted capacitance increased linearly with the hydrogel area, confirming that the measured response arises from the geometrical capacitance of the ionic conductor/electrode interface rather than from an isolated device-specific contribution. The capacitance density showed only a weak dependence on frequency in the 1–100 Hz range, decreasing slightly from approximately $9.6 \times 10^{-3}$ F $m^{-2}$ at 1 Hz to approximately $8.8 \times 10^{-3}$ F $m^{-2}$ at 100 Hz. This weak frequency dependence indicates that the mobile ions in the hydrogel can follow the applied electric field on the timescale relevant for the transistor measurements. This behavior is consistent with electric-double-layer formation, in which mobile ions accumulate at the gate/electrolyte and electrolyte/semiconductor interfaces, providing strong electrostatic coupling between the gate and the channel[12,14].

The measured areal capacitance, on the order of $9 \times 10^{-3}$ F $m^{-2}$, is sufficiently high to enable low-voltage field-effect operation of $MoS_2$ devices. Although commercially available TENS hydrogels are not specifically designed as transistor gate dielectrics, their water-rich polymer network and mobile ionic content allow them to function as an ionic-gating medium, similar in concept to polymer electrolytes, ionic liquids, and ion gels previously used in electrolyte-gated transistors[11,12,15]. Importantly, unlike laboratory-prepared electrolytes, these hydrogels can be used directly without synthesis, solvent handling, curing steps, or glovebox processing. This makes them particularly attractive for teaching laboratories, where simplicity, robustness, and safe handling are more important than optimizing the ultimate device performance.

After establishing the electrical double-layer behavior and effective capacitance of the commercially available ionic hydrogel (Figure 1), we next demonstrate its practical use as a gate dielectric for field-effect transistor operation. Having confirmed that the material provides sufficiently large interfacial capacitance and stable electrostatic modulation, we proceed to fabricate simple ionic-gel-gated devices using only off-the-shelf components and minimal processing steps.

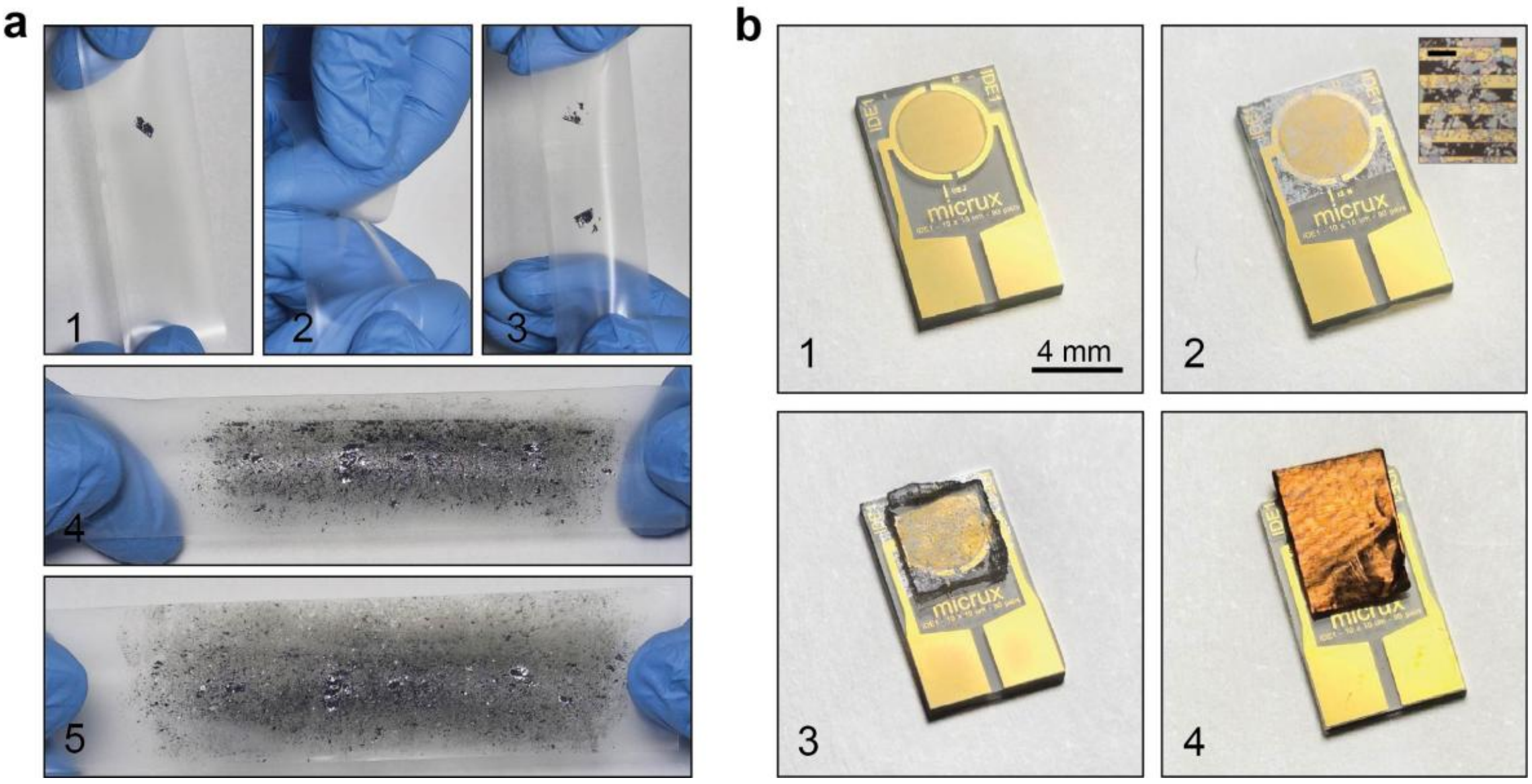

**Figure 2 | Step-by-step fabrication of ionic-gel-gated field-effect transistors using off-the-shelf materials. (a)** High-throughput mechanical exfoliation of $MoS_2$ using adhesive Nitto SPV 224 tape, shown here in a manual implementation of our recently developed roll-to-roll mechanical exfoliation approach. Panels 1–3 illustrate successive exfoliation steps, while panels 4 and 5 show the same tape after 50 and 100 exfoliation cycles, respectively, evidencing the accumulation of exfoliated material and the scalability of the process. **(b)** Assembly steps of the transistor device. Panel 1 shows the bare commercial interdigitated electrode (IDE) substrate. Panel 2 shows the device after transferring $MoS_2$ flakes onto the electrodes via thermal-release transfer from the exfoliation tape. The inset shows a magnified view of the channel. The scale bar represents 20 µm. Panel 3 shows the device after placing a commercially available ionic hydrogel as the gate dielectric. Panel 4 shows the final device after adding a copper tape serving as the top gate contact.

The complete device fabrication workflow is illustrated in Figure 2. The process begins with the preparation of the semiconducting channel material via high-throughput mechanical exfoliation, implemented here in a manual format inspired by our previously developed roll-to-roll exfoliation approach (Figure 2a)[16–18]. First a small piece of bulk $MoS_2$ crystal is placed on the adhesive side of a Nitto SPV 224 tape. Then the tape is folded onto itself to sandwich the crystal between the adhesive surfaces of the tape by gently pressing it. The tape is then peeled apart which allows thinner flakes to be isolated from the bulk source breaking the weak van der Waals forces between the layers. By progressively repeating this peeling process, we distribute thin flakes over the adhesive surface. After exfoliation step, a small cut piece of Nitto tape containing $MoS_2$ flakes is brought into contact with the active electrode region of commercially available interdigitated electrode (IDE) substrates (Micrux). Gentle pressure was applied with a cotton swab to improve contact between the flakes and the electrode surface. These IDE substrates consist of multiple gold electrode fingers with a width of 10 μm which are separated by 10 μm gaps (distance between adjacent fingers). Figure S1 gives a representative illustration of the channel geometry and the corresponding estimation of the channel width. After attaching the flake containing tape, the sample is heated on a hot plate at 100 °C for 5 minutes, after which the tape is peeled off, leaving $MoS_2$ flakes on the substrate, bridging the source and drain electrodes. Two successive $MoS_2$ flake transfer steps from the Nitto tape to the substrate were performed. After flake transfer, a small piece of ionic hydrogel, previously characterized in Figure 1, is subsequently placed in contact with the $MoS_2$ covered region to act as the gate electrolyte, and a metallic top contact (copper tape or aluminum foil) is added to complete the gate electrode (Figure 2b). We refer the reader to Video S1, which shows the complete fabrication process of an ion-gel-gated transistor.

Pre-patterned IDEs can be used directly as source and drain contacts without any additional lithography or metal deposition step. They are widely used in electrochemistry, biosensing, and impedance-based sensing applications, which have led to their broad commercial availability in standardized formats from multiple suppliers. As a result, these prepatterned electrode platforms are relatively inexpensive (typically below 10 € per device), making them particularly well suited for teaching laboratories and student practicums. This straightforward fabrication sequence enables the realization of functional ionic-gel-gated field-effect transistors without cleanroom processing, custom dielectric preparation, or specialized handling environments, making the approach well suited for teaching laboratories and rapid prototyping.

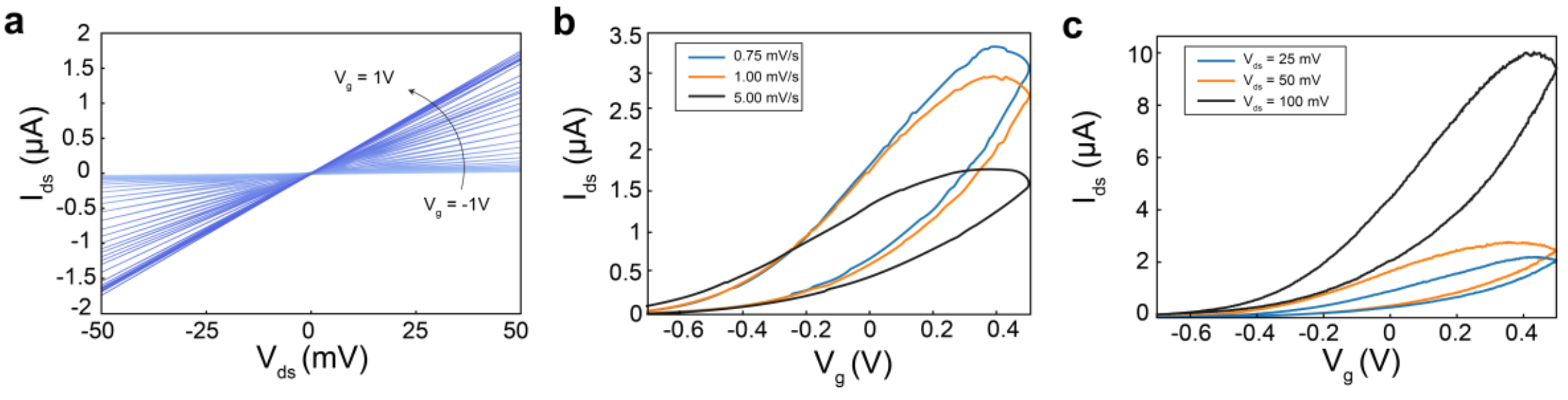

**Figure 3. Electrical characterization of ionic-gel-gated $MoS_2$ field-effect transistors. (a)** Output characteristics (source–drain current $I_{ds}$vs source–drain voltage $V_{ds}$) measured for gate voltages ranging from −1 V to +1 V, showing effective electrostatic modulation of channel conductance by the ionic gel gate. **(b)** Transfer characteristics ($I_{ds}$ vs gate voltage $V_g$) measured at $V_{ds}$ = 50 mV for different gate voltage sweep rates (0.75, 1.00, and 1.50 mV $s^{-1}$), showing pronounced hysteresis associated with slow ion migration and electric double-layer formation at the electrolyte–semiconductor interface. **(c)** Transfer characteristics measured at different $V_{sd}$ (25, 50, and 100 mV), illustrating the dependence of the device response on the applied drain field.

The electrical response of the fabricated devices was evaluated under ambient laboratory conditions using a homebuilt probe station. Representative measurements are shown in Figure 3. Electrical biasing and current measurements were performed using a Keithley 2450 source-measure unit (SMU), while the gate voltage was applied to the ionic hydrogel through a copper-tape gate electrode using two benchtop programmable power supplies connected back-to-back to enable bipolar voltage sweeps. It is important to note that, given the observed current levels lie in the microampere range, a lower-cost SMU instrument would be fully adequate for device characterization. Moreover, due to the millimeter-scale contact pads of the commercial interdigitated electrode substrates, electrical connections can be conveniently made using standard alligator clips, avoiding the need for micromanipulated probe needles if desired.

The output characteristics of the device (Figure 3a) show a progressive modulation of the drain current as the gate voltage is swept from negative to positive values, confirming that the commercial hydrogel can operate as an effective ionic gate. Positive gate bias increases the channel conductance, as expected for electron accumulation in $MoS_2$, whereas negative gate bias reduces the conductance. This behavior provides a direct classroom demonstration of the field effect: the channel current can be controlled electrostatically without changing the source–drain contact geometry or the semiconductor material itself. The source–drain voltage ($V_{ds}$) was intentionally restricted to a low range (-50 mV to + 50 mV) to maintain low-field operation and to minimize possible parasitic contributions from ionic conduction and electrochemical processes in the hydrogel contacting the source–drain region. Within this operating range, the systematic change in the $I_{ds}$ - $V_{ds}$ characteristics with gate voltage clearly demonstrates gate-controlled modulation of the channel conductance. Although these measurements do not access the saturation regime, they provide a direct experimental illustration of one of the central principles of field-effect transistor operation: electrostatic control of the channel conductance by the gate voltage.

The transfer characteristics further highlight the time-dependent nature of ionic gating. As shown in Figure 3b, the forward and reverse gate-voltage sweeps exhibit pronounced hysteresis, and the measured response depends on the gate-voltage sweep rate. Such hysteresis is not expected for an ideal, purely capacitive solid-state-gated transistor and, in the present device, is consistent with the slow redistribution of mobile ions in the hydrogel. When a gate voltage is applied, the mobile ions in the hydrogel redistribute and accumulate at the gate/electrolyte and electrolyte/semiconductor interfaces, where electric double layers are formed. Since ionic redistribution is not instantaneous, the interfacial ionic configuration, and consequently the carrier density in the $MoS_2$ channel, cannot immediately follow changes in $V_g$. Therefore, during the forward and reverse sweeps, the ionic distribution at the hydrogel/$MoS_2$ interface can be different at the same applied $V_g$, resulting in different values of $I_{ds}$ and giving rise to the observed hysteresis.

During a forward sweep from negative to positive gate voltage, the interfacial ionic distribution progressively reverses. The forward sweep starts at the most negative gate voltage, where cations and anions in the ionic gel are preferentially accumulated at the gate/electrolyte and semiconductor/electrolyte interfaces, respectively. As the gate voltage is increased toward more positive values, this ionic distribution gradually reverses as the cations and anions migrate toward the opposite interfaces. This redistribution modifies the interfacial electric double layers, which are formed by ionic charge in the hydrogel and oppositely charged electronic carriers in the adjacent gate electrode and semiconductor. Since $MoS_2$ exhibits n-type transport, the positive ionic charge accumulated at the hydrogel/$MoS_2$ interface electrostatically increases the electron carrier density in the channel. These additional mobile electrons contribute to charge transport between the source and drain, resulting in increased channel conductance and $I_{ds}$. Under negative gate bias, the ionic distribution is reversed, leading to electron depletion in the $MoS_2$ channel and a corresponding decrease in conductance.

The dependence on sweep rate provides additional evidence of the time-dependent ionic response. Lower gate-voltage sweep rates provide more time for the ions to redistribute during sweep. Consequently, during the positive sweep, larger gate-induced electron accumulation and higher $I_{ds}$ can be observed at lower sweep rates than at higher sweep rates. From an educational perspective, this sweep-rate dependence provides a simple experimental means of investigating the dynamics of ionic gating: by comparing transfer curves acquired at different sweep rates, students can directly observe how the timescale of ionic redistribution affects the transistor response and contributes to hysteresis.

The transfer curves measured at different drain biases (Figure 3c) show that the apparent gate response also depends on the applied source–drain voltage. Increasing the drain bias increases the measured drain current and can modify the shape of the transfer curve because the lateral electric field, contact resistance, and gate-induced carrier density all contribute to the measured two-terminal response. This provides an additional teaching opportunity: students can directly see that transistor characteristics are not determined only by the gate voltage, but also by the measurement conditions.

Although ionic gating introduces additional phenomena such as ion redistribution and hysteresis compared with conventional solid-state gating, it considerably simplifies the experimental implementation of the gate dielectric by eliminating dielectric deposition and other cleanroom-based processing steps typically required for conventional transistor fabrication. This experimental simplicity enables students to assemble and characterize a functioning device with minimal laboratory infrastructure. Together, the output and transfer measurements demonstrate the key operating principles of ionic-gel-gated $MoS_2$ field-effect transistors.

While the devices presented here were fabricated using the manual high-throughput exfoliation procedure described in Figure 2, the same measurement approach applied to devices prepared using automated roll-to-roll mechanical exfoliation, which yields improved film uniformity[16,17]. Electrical datasets from representative roll-to-roll-fabricated device are provided in the Supporting Information (Figure S2 and S3).

For readers interested in theoretical modelling, we refer to a study that implements finite-element modelling using COMSOL Multiphysics to investigate coupled transport in electric-double-layer-gated two-dimensional semiconductor transistors[19]. In addition, an introductory COMSOL model of an electrolyte-gated organic field-effect transistor is available as a practical example for readers interested in implementing this type of simulation[20].

## CONCLUSIONS

We have presented a simple and accessible experimental platform that enables the fabrication and characterization of functional field-effect transistors using commercially available materials and minimal laboratory infrastructure. By employing off-the-shelf hydrogel ionic conductors originally developed for transcutaneous electrical nerve stimulation (TENS) devices, we demonstrate that effective electrostatic gating of two-dimensional semiconductor channels can be achieved without the need for cleanroom processing, custom electrolyte synthesis, or controlled environments.

Electrical measurements confirm that the hydrogel forms a high-capacitance electric double layer capable of modulating channel conductance at low gate voltages, enabling clear field-effect behavior. The resulting devices can be fabricated using straightforward mechanical exfoliation and simple assembly steps and characterized with readily available instrumentation under ambient conditions. The millimeter-scale device geometry and microampere-level currents further simplify electrical interfacing, allowing practical implementations using low-cost measurement equipment and basic electrical connections.

Importantly, the fabrication and measurement workflow is compatible with both manual and automated material preparation methods. While the procedures demonstrated here prioritize simplicity and accessibility for instructional settings, higher uniformity and scalability can be achieved using roll-to-roll mechanical exfoliation, as illustrated in the Supporting Information.

Overall, this work demonstrates that hands-on fabrication and operation of modern transistor devices can be integrated into standard teaching laboratories without access to semiconductor foundries. The platform provides an accessible laboratory activity in which students can fabricate an emerging electronic device and subsequently investigate field-effect characteristics through direct electrical measurements. We anticipate that this platform will be broadly useful for education, rapid prototyping, and resource-limited research environments.

## MATERIALS AND METHODS

Thin-film interdigitated electrode substrates were supplied from Micrux Technologies (www.micruxfluidic.com). The $MoS_2$ flakes were obtained from manual high-throughput mechanical exfoliation of natural molybdenite mineral (Molly Hill Mine, Quebec, Canada). Commercially available conductive hydrogel electrodes (C100AE EMS hydrogel pads, model JP-EMS-11) used in transcutaneous electrical nerve stimulation (TENS) pads were used as ionic interface layer on the channel. The hydrogel served as the ionic conductor and gate dielectric. These hydrogels are water-rich polymer networks containing mobile ions, which provide ionic conductivity and conformal contact with the semiconductor channel.

### Roll-to-roll Mechanical Exfoliation

For comparison and supporting experiments, $MoS_2$ flakes obtained by roll-to-roll mechanical exfoliation were also used. This exfoliation method is based on the continuous exfoliation of bulk van der Waals crystals between two rolling cylinders wrapped with adhesive Nitto tape. The repeated rolling process enables the production of large amounts of exfoliated material over a larger area, improving the probability of forming continuous flake networks across the interdigitated electrode channels. For further details on the method, we refer the reader to our previous papers.[16,17]

### Hydrogel capacitance measurements

The capacitance of the commercial ionic hydrogel was characterized using a simple resistor–capacitor (RC) circuit. Square pieces of hydrogel with different lateral areas were cut and placed between two metallic electrodes, forming a parallel-plate capacitor geometry. The hydrogel capacitor was connected in series with a known resistor, and a square-wave pulses were applied to the circuit by a function generator.

The voltage response across the hydrogel capacitor was monitored using Moku:Go instrument (liquidinstruments.com), to obtain the charging and discharging transients. The characteristic RC time constant was determined from the charging transient as the time required for the capacitor voltage to reach 63% of its steady-state value. The capacitance was then calculated using $C = \tau/R$, where $\tau$ is the extracted time constant and R is the known series resistance. This procedure was repeated for hydrogel capacitors with different areas and at different excitation frequencies. The areal capacitance was obtained from the slope of the capacitance-versus-area plot.

### Electrical measurements

All electrical measurements were performed under atmospheric pressure at room temperature using a home-built probe station. The source–drain terminals of the interdigitated electrode devices were biased using probes connected to a Keithley 2450 source-meter unit. The gate voltage was applied to the copper gate electrode through the ionic hydrogel using two programmable benchtop power supplies connected in a back-to-back configuration, enabling bipolar gate-voltage sweeps.

Output characteristics were recorded by sweeping the drain–source voltage while keeping the gate voltage fixed at selected values. Transfer characteristics were measured by sweeping the gate voltage while maintaining a constant drain–source bias. Different gate-voltage sweep rates and drain-bias values were used to evaluate the hysteretic response of the ionic-gel-gated devices and the effect of electric-double-layer formation at the hydrogel/$MoS_2$ interface.

## ASSOCIATED CONTENT

### Supporting Information

Supplementary Information includes:

**Figure S1.** Channel-width estimation for the circular multichannel electrode geometry.

**Figure S2.** Electrical characterization of a roll-to-roll exfoliated $MoS_2$ device with ionic-gel gating.

**Figure S3.** Extended gate-voltage operation of a roll-to-roll exfoliated another $MoS_2$ transistor.

## AUTHOR INFORMATION

### Corresponding Authors

Andres Castellanos-Gomez andres.castellanos@csic.es

### Author Contributions

Y.S. and T.K. contributed equally to this work. Y.S. and T.K. led data curation, formal analysis, investigation, methodology, visualization, and original draft writing. J.J.R supported supervision and methodology. A.C.-G. led conceptualization, funding acquisition, project administration, resources, and supervision, supported the methodology and original draft writing. All authors contributed to review and editing and approved the final version of the manuscript.

## ACKNOWLEDGEMENTS

We acknowledge the Severo Ochoa Centers of Excellence Program through the grant CEX2024-001445-S. A.C-G. acknowledges funding by the Ministry of Science, Innovation and Universities (Spain) through the projects PDC2023-145920-I00 and PID2023-151946OB-I00 and the European Research Council through the ERC-2024-PoC StEnSo (grant agreement 101185235) and the ERC-2024-SyG SKIN2DTRONICS (grant agreement 101167218). Y.S. acknowledges funding by the Ministry of Science, Innovation and Universities (Spain) through the grant PRE2021-098348. J.J.R. hired under the Generation D initiative, promoted by Red.es, an organisation attached to the Ministry for Digital Transformation and the Civil Service, for the attraction and retention of talent through grants and training contracts, financed by the Recovery, Transformation and Resilience Plan through the European Union's Next Generation funds.

# Supporting Information:

## Hands-on fabrication of 2D material field-effect transistors for the classroom using off-the-shelf ionic gel

*Yigit Sozen, Tayfun Kutlu, Juan J. Riquelme, Andres Castellanos-Gomez**

*2D Foundry research group. Instituto de Ciencia de Materiales de Madrid (ICMM-CSIC), Madrid, E-28049, Spain.*

*corresponding author andres.castellanos@csic.es

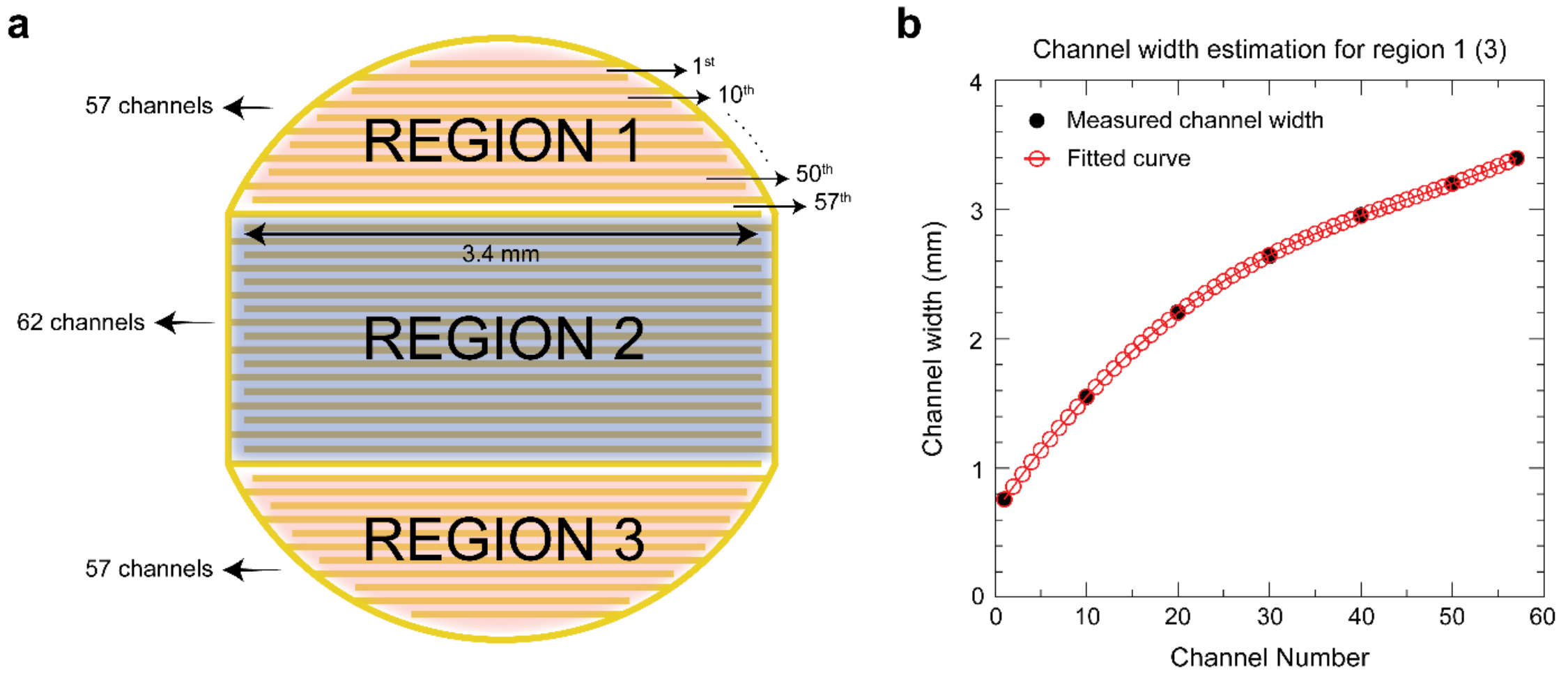

**Figure S1 | Channel-width estimation for the circular multichannel electrode geometry.**
**(a)** Schematic representation of the circular multichannel electrode layout divided into three regions according to the local channel geometry. Region 1 and Region 3 each contain 57 channels, while Region 2 contains 62 channels. The central rectangular region has a fixed width of 3.4 mm, whereas the effective channel width in the upper and lower curved regions varies due to the circular boundary of the pattern. Representative channel positions, including the 1st, 10th, 50th, and 57th channels, are indicated. **(b)** Channel-width estimation for Region 1 as a function of channel number. Black symbols correspond to the measured channel widths, while red open circles represent the fitted curve used to interpolate the effective width of each channel across the curved region. The fitted channel-width profile accounts for the gradual increase in channel length from the edge toward the center of the circular electrode pattern.

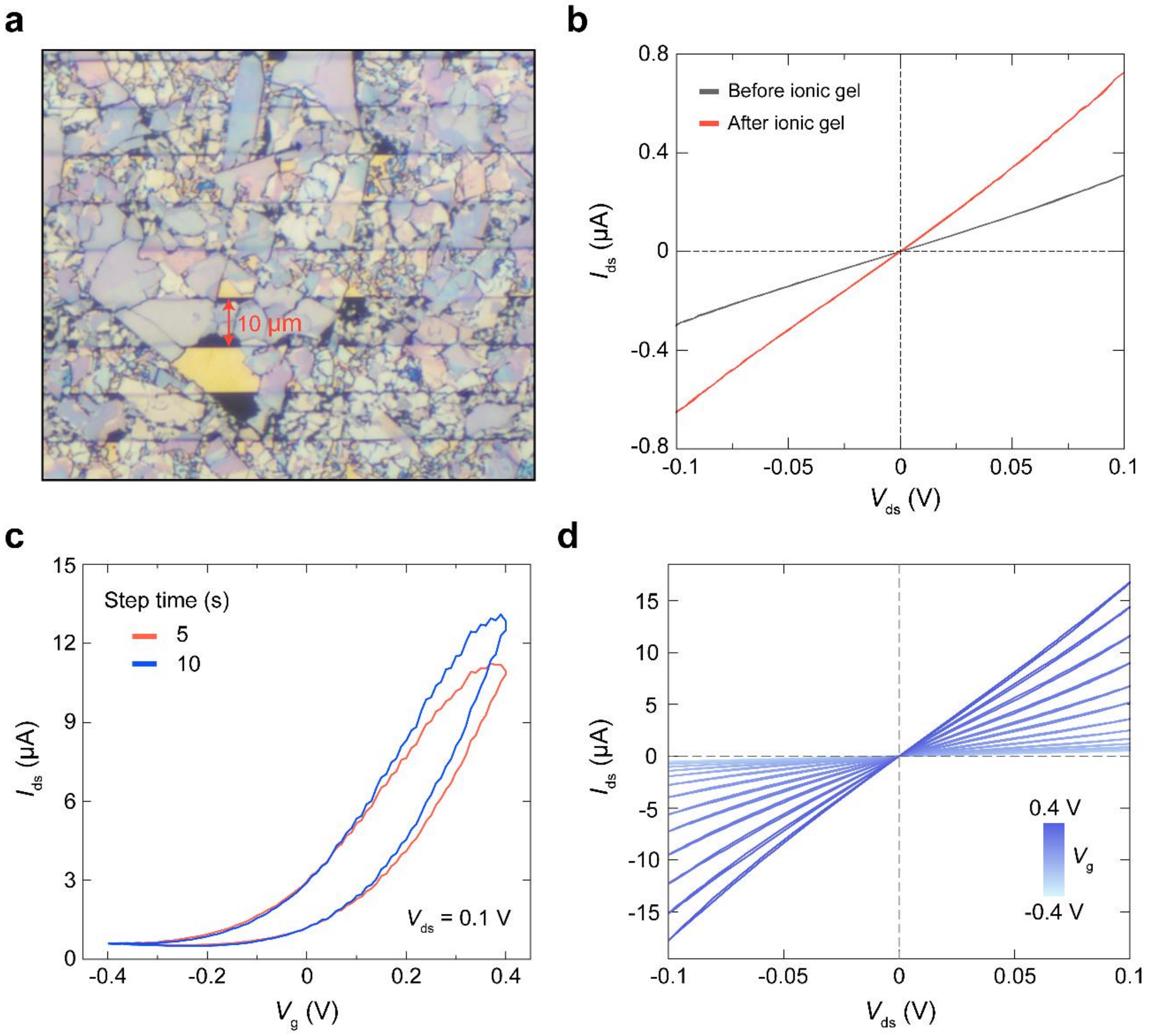

**Figure S2 | Electrical characterization of a roll-to-roll exfoliated $MoS_2$ device with ionic-gel gating.** (a) Optical microscopy image of a $MoS_2$ flake obtained by roll-to-roll exfoliation and integrated into a two-terminal device. The channel length between the source and drain electrodes is approximately 10 μm. (b) $I_{ds}$-$V_{ds}$ characteristics measured before and after ionic-gel deposition. The increased source–drain current after gel integration indicates enhanced channel conductance, consistent with electrostatic doping or improved charge accumulation induced by the ionic medium. (c) Transfer characteristics measured at $V_{ds}$ = 0.1 V for two dwell times of 5 and 10 s, with the gate voltage swept in 0.01 V increments. The hysteretic gate response reflects the slow ionic redistribution and electric-double-layer formation in the ionic gel. (d) Gate-dependent output characteristics recorded for gate voltages from −0.4 to 0.4 V.

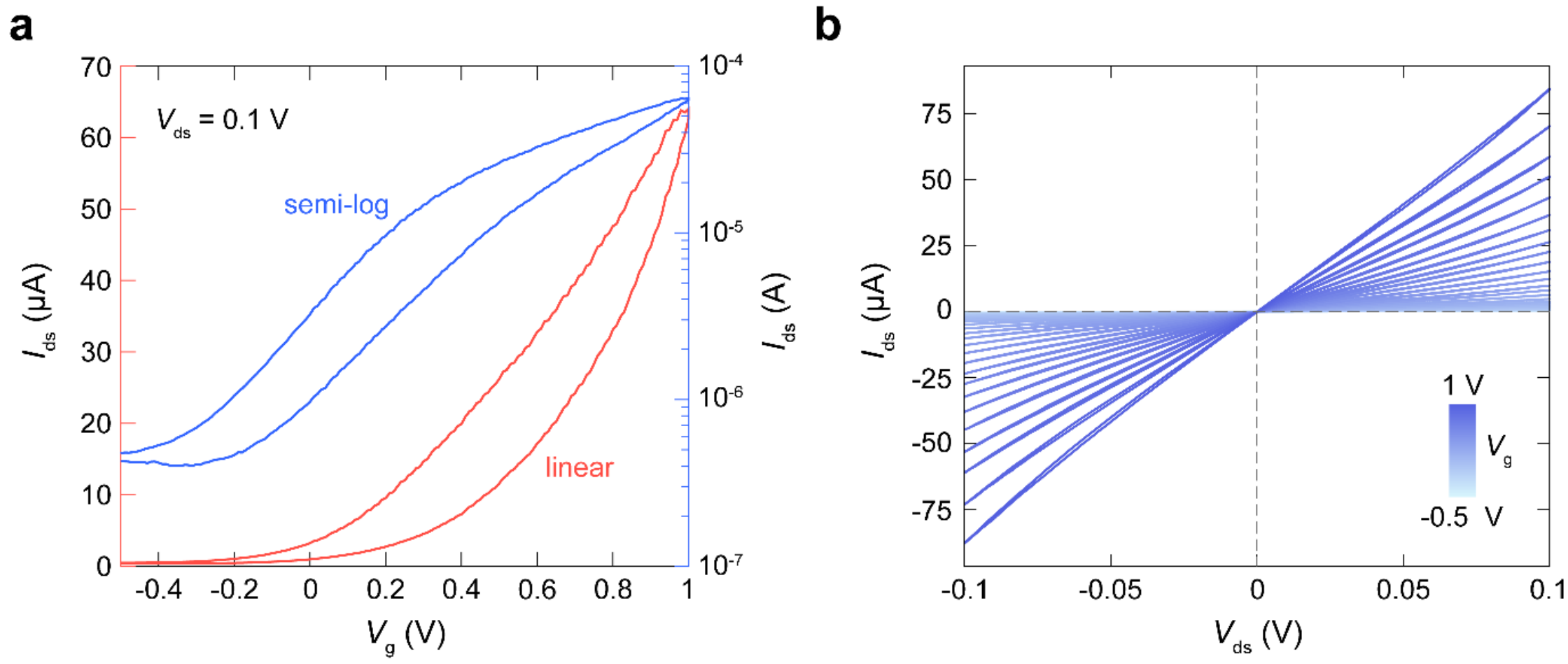

**Figure S3 | Extended gate-voltage operation of a roll-to-roll exfoliated another $MoS_2$ transistor.** (a) Transfer characteristics of an ionic-gel gated $MoS_2$ transistor fabricated from roll-to-roll exfoliated flakes and measured at $V_{ds}$ = 0.1 V. The same dataset is shown on linear and semi-logarithmic current scales to illustrate two complementary ways of evaluating transistor behavior: the linear plot emphasizes the absolute current increase, whereas the semi-logarithmic plot highlights the relative gate-induced modulation and off-state response. Compared with the roll-to-roll exfoliated device shown in Figure S2, this device was measured over an extended gate-voltage window up to $V_g$ = 1 V, allowing the n-type field-effect response to be more clearly identified. (b) Gate-dependent output characteristics recorded from $V_g$ = -0.5 to 1 V.